\documentclass[12pt,a4paper]{article}
\usepackage{amsmath}
\usepackage{amsxtra}
    \usepackage{amstext}
    \usepackage{amssymb}
    \usepackage{latexsym}
    \usepackage{graphicx}
\usepackage{color}
\usepackage{graphics}

\newcommand{\p}{\vspace{6pt}\noindent}

\advance\textheight by \topskip
\makeatletter

\@addtoreset{equation}{section}
\def\section{\@startsection {section}{1}{\z@}{-8.5ex plus -1ex minus
 -.2ex}{3.3ex plus .2ex}{\large\bf}}
\def\subsection{\@startsection{subsection}{2}{\z@}{-3.25ex plus
 -1ex minus -.2ex}{1.5ex plus .2ex}{\bf}}
\def\subsubsection{\@startsection{subsubsection}{3}{\z@}{-3.25ex plus%
 -1ex minus -.2ex}{1.5ex plus .2ex}{\sl}}

\begin{document}

\begin{titlepage}
\vspace*{-2cm}
\begin{flushright}
\end{flushright}

\vspace{0.3cm}

\begin{center}
{\Large {\bf Infinite dimension reflection matrices in
the\\\vspace{10pt}
sine-Gordon model with a boundary}} \vspace{1cm} {\Large {\bf }}\\
\vspace{1cm} {\large  E.\ Corrigan\footnote{\noindent E-mail: {\tt
edward.corrigan@york.ac.uk}}}\\
\vspace{0.5cm}
{\em Department of Mathematics \\ University of York, York YO10 5DD, U.K.} \\
\vspace{0.3cm} {\large and}\\ \vspace{0.5cm}
{\large C.\ Zambon\footnote{\noindent E-mail: {\tt cristina.zambon@durham.ac.uk}}} \\
\vspace{0.3cm}
{\em Department of Physics \\ Durham University, Durham DH1 3LE, U.K.} \\

\vspace{2cm} {\bf{ABSTRACT}}\\ \end{center} \p Using the sine-Gordon
model as the prime example an alternative approach to  integrable
boundary conditions for a theory restricted to a half-line is
proposed. The main idea is to explore the consequences of taking
into account the topological charge residing on the boundary and the
fact it changes as solitons in the bulk reflect from the boundary.
In this context, reflection matrices are intrinsically infinite
dimensional, more general than the two-parameter
Ghoshal-Zamolodchikov reflection matrix, and related in an intimate
manner with defects.  \vspace{.5cm}

\p \\

\vfill
\end{titlepage}

\section{Introduction}

\p The investigation of two-dimensional classical and quantum
integrable field theories with one or two boundaries is now several
decades old (see for example
\cite{cherednik1984}-\cite{Bajnok:2002vp}), yet it is incomplete
with many interesting outstanding questions remaining to be
answered, even in the context of the much-studied affine Toda
models. More recently, there has been some interest in the
definition and properties of integrable defects (see, for example
\cite{delfino94}-\cite{DoikouAvan11}), which is really the study of
allowable discontinuities, in the sense of permitting field
discontinuities that do not upset integrability, and already there
have been suggestions \cite{Bajnok:2004jd, bajnok07,bu2008}
concerning possible relationships between the two ideas. The purpose
of this article is to add to these suggestions having in mind models
that allow solitons and using the sine-Gordon model as the prime
example.

\p Basically, a defect is a bulk phenomenon but placing a defect
close to, or even at, a boundary will effectively modify the
reflection matrix associated with the boundary in several ways.
First, the defect is able to introduce additional parameters, and
second, even if in the absence of the defect a reflection matrix
contained no explicit dependence on the topological charge residing
at the boundary  - since boundary conditions only exceptionally
preserve topological charges carried by solitons - the defect will
(almost) inevitably introduce dependence on topological charge in
the modified reflection matrix. So, it would seem sensible to
reconsider the boundary Yang-Baxter equation, or reflection
equation, allowing the boundary to carry topological charge, and the
reflection matrix to be infinite-dimensional, then attempt to
classify its solutions.

\p If the sine-Gordon model is considered as the primary example, on
the grounds that it is the simplest of the affine Toda field
theories, one might expect the solutions to the reflection equation
to be one of three types: (1) the coefficients of the reflection
matrix could be independent of topological charge (basically, this
is the Ghoshal-Zamolodchikov solution \cite{gz1994}, though
reformulated slightly); (2) the reflection matrix is related to a
Ghoshal-Zamolodchikov solution modified by a defect; or (3) neither
of these - in which case the reflection matrix may describe a new
type of boundary condition, which might turn out to be a defect
`fused' with a boundary. The latter idea might appear strange at
first sight yet it is worth recalling that defects themselves may be
fused together and the essential difference between an unfused and a
fused pair is the evident fact that a soliton can propagate in the
gap between the unfused pair and there is no gap between a fused
pair \cite{cz2010}. One need look no further than a Neumann boundary
condition (placing the boundary at $x=0$), $u_x(0,t)=0$, to realise
that a boundary can store an unlimited quantity of topological
charge since $N$ solitons approaching the boundary eventually
reflect as $N$ anti-solitons, or vice versa, with the boundary
picking up 2$N$ units of charge. In effect, the boundary value of
$u$ changes by $4\pi N$ during the process. On the other hand, a
Dirichlet boundary condition, for example $u(0,t)=u_0$, ensures
solitons reflect as solitons and the boundary charge does not change
during the process.

\p Under reasonably general assumptions, effectively meaning the
absence of any dynamical variables confined to the boundary,
classically integrable boundary conditions have been determined
\cite{cdrs1994} for any of the (real) affine Toda field theories in
the following form
\begin{equation}\label{todaboundary}
 u_x(0,t)=-\frac{\partial{\cal B}}{\partial u},\quad {\cal B}
 =\sum_{r=0}^n b_r\, e^{\alpha_r\cdot u(0,t)/2},
\end{equation}
where  $u$ is a multicomponent field, the vectors $\alpha_r$ are the
Euclidean parts of the affine roots described by any of the
Dynkin-Kac diagrams \cite{kacbook}, the coefficients $b_r$ are
constants. These boundary conditions possess curious properties
\cite{bcdr1995}. With a single exception, the most unexpected
concerns the affine Toda field theories based on $a_n^{(1)},\
d_n^{(1)},\ e_n^{(1)}$ root data. Apart from the sine-Gordon model
(the exception, based on $a_1^{(1)}$ data), which has two free
parameters $b_1,b_2$ associated with a general integrable boundary
condition, these models do not appear to allow any free parameters
associated with a boundary. Instead, there is a discrete set of
non-zero values for the coefficients $b_r$ amounting to a set of
sign choices (Neumann is always a possibility with $b_r=0$ for all
$r$). The other models (based on non-simply-laced data) fare a
little better, in the sense that there are usually one or two free
parameters but never as many as the rank of the associated affine
algebra. In the complex affine Toda models (where there are complex
solitons), there are other boundary conditions \cite{delius1998} of
the form
\begin{equation}\label{todadirichlet}
{\rm Im}\,u(0,t)=2\pi\lambda,\quad {\rm Re} \,u_x(0,t)=0,
\end{equation}
where $\lambda$ is a weight. Clearly these are a mixture of
Dirichlet and Neumann conditions, and they are soliton-preserving.
In the complex models the conditions \eqref{todaboundary} are not
soliton-preserving and will inevitably change the topological charge
on the boundary. It is possible that the assumptions made to reach
these results are too stringent but it has not yet been found how to
relax them. It is also possible that dressing a boundary with one or
more defects may add additional parameters and provide a clue
towards the goal of generalising the boundary conditions. On the
other hand, this is easier said than done because, so far, defects
are only established for the affine Toda models based on
$a_n^{(1)},\ n\ge 1$ and $ a_2^{(2)}$ \cite{cz2009,cz2011}.

\section{Generalised reflection matrices}

\p Consider the sine-Gordon model, a typical reflection matrix is
\begin{equation}\label{reflectionmatrix}
 R_{a\,\alpha}^{b\,\beta}(\theta),
\end{equation}
where the labels $a,b=\pm 1$ (or simply $\pm$) represent the incoming and outgoing soliton, the labels $\alpha,\beta$ represent the initial and final topological charge carried by the boundary, and $\theta$ is the rapidity of the incoming soliton. In much of the earlier literature the boundary topological charge labels are simply suppressed but here it will be useful to keep them explicit. Clearly, topological charge conservation requires
\begin{equation}
 a+\alpha = b+\beta,
\end{equation}
and the boundary labels must be either even or odd since they may only change by $\pm 2$. The reflection matrix will also depend upon the bulk coupling and boundary parameters. In the more general affine Toda models, both sets of labels should be replaced by suitable sets of weights.

\p The reflection equation, or boundary Yang-Baxter equation \cite{cherednik1984}, is a condition that ensures compatibility with the bulk soliton S-matrix. Thus
\begin{equation}\label{bybe}
 R_{a\,\alpha}^{q\,\beta}(\theta_a)\, S_{b\, q}^{p\,s}(\theta_b+\theta_a) R_{p\,\beta}^{r\,\gamma}(\theta_b)\,S_{s\,r}^{d\,c}(\theta_b-\theta_a)=S_{b\, a}^{p\,q}(\theta_b-\theta_a)R_{p,\alpha}^{r\,\beta}(\theta_b)S_{q\, r}^{s\, c}(\theta_a+\theta_b)R_{s\,\beta}^{d\,\gamma}(\theta_a),
\end{equation}
where repeated indices are summed. In the sine-Gordon model, the bulk S-matrix \cite{ZZ78} is given by a $4\times 4$ matrix depending on the parameter $\Theta\equiv\Theta_{12}=(\theta_1-\theta_2)$ whose elements different from zero are:
\begin{eqnarray}\label{Smatrix}
S^{+\,+}_{+\,+}(\Theta)&=&S^{-\, -}_{-\,-}(\Theta)=\left(\frac{qx_1}{x_2}-\frac{x_2}{qx_1}\right)\rho_s(\Theta)\equiv a(\Theta)\, \rho_s(\Theta),\nonumber \\
S^{-\,+}_{+-}(\Theta)&=&S^{+\,-}_{-\,+}(\Theta)
=\left(\frac{x_1}{x_2}-\frac{x_2}{x_1}\right)\rho_s(\Theta)\equiv b(\Theta)\,\rho_s(\Theta),\nonumber \\ S^{+\,-}_{+\,-}(\Theta)&=&S^{-\,+}_{-\,+}(\Theta)
=\left(q-\frac{1}{q}\right)\rho_s(\Theta)\equiv c\,\rho_s(\Theta),
\end{eqnarray}
with
$$x_p=e^{\gamma\theta_p},\quad q=-e^{-i\pi\gamma}=e^{-4i\pi^2/\beta^2},\quad \gamma=\frac{4\pi}{\beta^2}-1.$$
The multiplicative factor $\rho_s(\Theta)$ is given by:
\begin{equation}\label{rhos}
\rho_S(\Theta)=\frac{\Gamma(1-z-\gamma)\Gamma(1+z)}{2\pi i}\,\prod^{\infty}_{k=1}R_k(\Theta)R_k(i\pi-\Theta),\quad z=\frac{i\gamma\Theta}{\pi},
\end{equation}
with
$$R_k(\Theta)=\frac{\Gamma(z+2k\gamma)\Gamma(1+z+2k\gamma)}
{\Gamma(z+(2k+1)\gamma)\Gamma(1+z+(2k-1)\gamma)}.$$

\p There are many solutions to \eqref{bybe} that may be written
\begin{equation}\label{Rmatrix}
 R_{a\,\alpha}^{b\,\beta}(\theta)=\left(\begin{array}{ll}
                                         r_+(\alpha,x)\,\delta_\alpha^\beta& s_+(\alpha,x)\,\delta_\alpha^{\beta-2}\\
                                         s_-(\alpha,x)\,\delta_\alpha^{\beta+2}&r_-(\alpha,x)\,\delta_\alpha^\beta\\
                                        \end{array}\right),
\end{equation}
where the reflection factor is written in block form, and it is understood the blocks are labeled by the solitons. Typically, the coefficients appearing in \eqref{Rmatrix} will be functions of $\alpha$, the initial charge on the boundary, besides the rapidity, bulk coupling and  boundary parameters. However, there is a particular two-parameter class of solutions, referred to as the Ghoshal-Zamolodchikov solution, where there is no dependence on $\alpha$ other than via the Kronecker deltas. In this case the latter are there merely to keep track of the topological charge. This solution has the following form,
\begin{equation}\label{GZsolution}
  R_{a\,\alpha}^{b\,\beta}(\theta)=\sigma(\theta)\left(\begin{array}{cc}
                                         \left(r_1x+r_2/x\right)\delta_\alpha^\beta& k_0\left(x^2-1/x^2\right)\delta_\alpha^{\beta-2}\\
                                         l_0\left(x^2-1/x^2\right)\delta_\alpha^{\beta+2}&\left(r_2x +r_1/x\right)\delta_\alpha^\beta\\
                                        \end{array}\right),
\end{equation}
where $r_1,r_2,k_0,l_0$ are constants and $x=e^{\gamma\theta}$.  Without losing generality one may impose $r_1r_2=1,\ k_0=l_0$, by removing an overall constant factor and using a similarity transformation. The requirements of crossing and unitarity  restrict the overall scalar factor $\sigma(\theta)$ \cite{gz1994}.

\p A more general solution to \eqref{bybe} can be obtained quite easily by assuming the elements $s_\pm(\alpha,x)$ remain proportional to $x^2-1/x^2$ and the diagonal elements are proportional to a cubic in $x$ and $1/x$. Explicitly, apart from an overall function of rapidity, the solution is:
\begin{eqnarray}\label{cubicR}
\nonumber&&\ \ \ \ \ \ \ \ \ \ \ \ \ \ r_+(\alpha,x)=\left(x^2-1/x^2\right)\left(r_3q^{\alpha+1} x - r_4q^{-\alpha-1}/x\right) +r_1 x +r_2/x,\\
\nonumber &&\ \ \ \  \ \ \ \ \ \ \ \ \ r_-(\alpha,x)=\left(x^2-1/x^2\right)\left(r_4q^{-\alpha+1} x - r_3q^{\alpha-1}/x\right) +r_2x +r_1/x,\\
\nonumber &&s_+(\alpha,x)=\left(x^2-1/x^2\right)(k_0+k_1q^\alpha +k_2q^{-\alpha}),\quad s_-(\alpha,x)=\left(x^2-1/x^2\right)(l_0+l_1q^\alpha +l_2 q^{-\alpha})\\
&&\ \ \ \ \ \   k_1l_1=-r_3^2,\quad k_2l_2=-r_4^2,\quad k_1 l_0 +q^2 k_0l_1=qr_2r_3,\quad k_0l_2 +q^2k_2l_0=qr_1r_4.
 \end{eqnarray}
Clearly, the expression \eqref{GZsolution} lies within this class of solution and is obtained by setting $r_3=r_4=0=k_1=k_2=l_1=l_2$. Another interesting special case concerns the choice $r_1=r_2=0$ for which the quantity $(x^2-1/x^2)$ becomes an overall factor. Actually, assuming all terms in \eqref{Rmatrix} are polynomial in $x$, it is not difficult to show that the polynomials must have common zeros, except for the roots of $x^4=1$, which may be present in the off-diagonal terms without also being roots of the diagonal terms. For this reason, the solution given above is general. More detail concerning these solutions is supplied in the appendix.

\p If a defect is placed before the boundary this will modify a
reflection factor  though not merely by altering its dependence on
rapidity. Since a defect also carries topological charge a new set
of labels will be added to the modified reflection matrix
\cite{sklyanin88, Bajnok:2004jd}. Thus, in detail,
\begin{equation}\label{TRT}
R_{a\,\alpha\,\tilde\alpha}^{b\,\beta\,\tilde\beta}(\theta)= T_{a\,\tilde\alpha}^{c\,\tilde\gamma}(\theta)R_{c\,\alpha}^{d\,\beta}(\theta)
\hat T_{d\,\tilde\gamma}^{b\,\tilde\beta}(\theta),
\end{equation}
where $\tilde\alpha,\tilde\beta,\tilde\gamma$ are topological charges associated with the defect, $T$ represents the transmission matrix for the soliton approaching the boundary, while $\hat T$ represents the transmission matrix for the soliton after reflection from the boundary. In fact $\hat T(\theta)$ is the inverse of $T(-\theta)$. The combination on the right hand side of \eqref{TRT} automatically satisfies a suitable generalisation of the compatibility relations \eqref{bybe}, provided the transmission matrix is also compatible with the bulk soliton S-matrix. The latter requires,
\begin{equation}\label{STT}
S_{a\,b}^{mn}(\theta_a-\theta_b)\,T{_{n\alpha}^{t\beta}}(\theta_a)\,
T{_{m\beta}^{s\gamma}}(\theta_b)=T{_{b\,\alpha}^{n\beta}}(\theta_b)\,
T{_{a\,\beta}^{m\gamma}}(\theta_a)\,S_{mn}^{st}(\theta_a-\theta_b).
\end{equation}
If the reflection matrix on the right hand side of \eqref{TRT}  is
proportional to $\delta_\alpha^\beta$ then the new reflection matrix
on the  left hand side is also proportional to $\delta_\alpha^\beta$
and satisfies precisely \eqref{bybe}.

\p
 In the sine-Gordon model solutions to \eqref{STT} are known and may therefore be used to construct solutions to \eqref{bybe}. Owing to topological charge conservation, they all possess the general shape
 \begin{equation}\label{Tg}
T^{b\,\beta}_{a\, \alpha}(\theta)=\rho(\theta)\left(\begin{array}{ll}
  a(\alpha, x) \,\delta_{\alpha}^{\beta} &
    b(\alpha, x) \,\delta_{\alpha}^{\beta-2} \\
    c(\alpha, x) \,\delta_{\alpha}^{\beta+2} &
   d(\alpha, x) \,\delta_{\alpha}^{\beta}\\
  \end{array}
\right).
\end{equation}
 In \cite{cz2010}, the Konik-LeClair solution \cite{konik97}  to \eqref{STT} was generalised to
\begin{equation}\label{Tmatrixgg}
T^{b \,\beta}_{a\, \alpha}(\theta)=\rho(\theta)\left(
  \begin{array}{cc}
    (a_+q^{-\alpha/2}x^{-1}+a_-q^{\alpha/2}x )\,\delta_{\alpha}^{\beta}& \mu(\alpha) \,\delta_{\alpha}^{\beta-2}\\
    \lambda(\alpha)\,\delta_{\alpha}^{\beta+2} & (d_+q^{-\alpha/2}x+d_-q^{\alpha/2}x^{-1})\,\delta_{\alpha}^{\beta} \\
  \end{array}
\right),
\end{equation}
with
$$\mu(\alpha)\,\lambda(\alpha+2)-\mu(\alpha-2)\,\lambda(\alpha)
=(q-q^{-1})\,(a_-d_-q^{\alpha}-a_+d_+q^{-\alpha}),$$
which implies
\begin{equation}
\label{constraintlamu}
\mu(\alpha-2)\,\lambda(\alpha)=a_-d_-q^{\alpha-1}+a_+d_+q^{-\alpha+1}+\gamma.
\end{equation}
Actually, it has not proved possible to generalise \eqref{Tmatrixgg}
further and it is most probably already the general solution (see
also \cite{Weston:2010cc}). The constraint \eqref{constraintlamu} is
satisfied by choosing, for instance,
$$\mu(\alpha)=b_+q^{-\alpha/2}+b_-q^{\alpha/2},\quad
\lambda(\alpha)=c_+q^{-\alpha/2}+c_-q^{\alpha/2},\quad
a_\pm\, d_\pm - b_\pm \,c_\pm=0,$$
where $a_\pm$, $b_\pm$, $c_\pm$, $d_\pm$ are otherwise free (complex) constants.
Then, setting $Q=q^{-\alpha/2}$, the expression \eqref{Tmatrixgg} becomes
\begin{eqnarray}\label{Tmatrixg}
T^{b \,\beta}_{a\, \alpha}(\theta)=
\rho(\theta)
\left(
\begin{array}{ccc}
  (a_+Q^{\alpha}x^{-1}+a_-Q^{-\alpha}x)\,\delta^{\beta}_{\alpha} &
  (b_+Q^{\alpha}+b_-Q^{-\alpha})\,\delta^{\beta-2}_{\alpha} \\
  (c_+Q^{\alpha}+c_-Q^{-\alpha} )\,\delta^{\beta+2}_{\alpha} & ( d_+Q^{\alpha}x+d_-Q^{-\alpha}x^{-1} )\,\delta^{\beta}_{\alpha} \\
\end{array}
\right).
\nonumber\\
&
\end{eqnarray}
Its inverse is given by:
\begin{equation}\label{Tmatrixginverse}
\left(T^{-1}\right)^{b \,\beta}_{a\, \alpha}=\frac{1}{\Delta\rho}\,\left(
\begin{array}{cc}
  ( d_+Q^{-\alpha-2}\,x+d_-Q^{\alpha+2}x^{-1} )\,\delta^{\beta}_{\alpha} &
 - (b_+Q^{-\alpha}+b_-Q^{\alpha})\,\delta^{\beta-2}_{\alpha} \\
 - (c_+Q^{-\alpha}+c_- Q^{\alpha})\,\delta^{\beta+2}_{\alpha} & (a_+Q^{-\alpha+2}\,x^{-1}+a_-Q^{\alpha-2}\, x)\,\delta^{\beta}_{\alpha} \\
\end{array}
\right),
\end{equation}
where $$\Delta=a_- d_+ q x^2+a_+d_-q^{-1}x^{-2}-b_+c_-q^{-1}-b_-c_+q.$$

\p Take the reflection matrix appearing on the right hand  side of
\eqref{TRT} to be diagonal, for example a Ghoshal-Zamolodchikov
solution \eqref{GZsolution} with $r_1=1/r_2\equiv r$, $k_0=l_0=0$,
corresponding to a Dirichlet boundary condition,
\begin{equation}\label{RmatrixDirichlet}
R_{c\,\alpha}^{d\,\beta}(\theta)=
\sigma(\theta)\left(
  \begin{array}{cc}
    (rx+x^{-1}r^{-1})\,\delta_\alpha^\beta & 0 \\
    0 & (rx^{-1}+xr^{-1})\,\delta_\alpha^\beta \\
  \end{array}
\right).
\end{equation}
Then, the result of the calculation \eqref{TRT} has the general
shape \eqref{Rmatrix} with coefficients of the following form (up to
an overall factor)
\begin{eqnarray}
&r_+({\alpha},x)=\left(x^2-{x^{-2}}\right)\,\left(a_-d_-q^{\alpha+1}r x
-a_+d_+q^{-\alpha-1}{r^{-1}x^{-1}}\right)
+x\,\left(g r-h{r^{-1}}\right)+
x^{-1}\,\left(g{r^{-1}}-h r\right),\nonumber\\
&r_-({\alpha,x})=\left(x^2-x^{-2}\right)\,\left(a_+d_+q^{-\alpha+1}r^{-1}x
-a_-d_-q^{\alpha-1}rx^{-1}\right)
+x\,\left(g{r^{-1}}-h r\right)+{x^{-1}}\,\left(g r-h{r^{-1}}\right),\nonumber\\
&s_+({\alpha},x)=\left(x^2-x^{-2}\right)\,\left(a_+b_-r^{-1}-a_-b_+ r
-a_-b_-q^{\alpha}r+a_+b_+q^{-\alpha}r^{-1}
\right),\nonumber\\
&s_-({\alpha},x)=\left(x^2-x^{-2}\right)\,\left(d_-c_+r-d_+c_-{r^{-1}}+d_-c_-q^{\alpha}r
-d_+c_+q^{-\alpha}r^{-1}\right),
\end{eqnarray}
with
$$g={a_-d_+}{q^{-1}}+a_+d_-q,\qquad h={b_-c_+}{q^{-1}}+b_+\,c_-q.$$

\p The Konik-LeClair \cite{konik97} (type I) solutions  are a
particular case of \eqref{Tmatrixg}. Set, for example,
$a_-=d_+=c_-=b_+=0$, multiply \eqref{Tmatrixg} by $x$ and collect an
overall factor. Then, \eqref{Tmatrixg} becomes
\begin{equation*}
T_{I}{^{\phantom{I}b \,\beta}_{\ a \alpha}}(\theta)=\rho_{I}(\theta)
\left(
 \begin{array}{cc}
   \nu^{-1/2}\, Q^{\alpha}\,\delta_{\alpha}^{\beta} &
    x\,\varepsilon\,Q^{-\alpha}\,\delta_{\alpha}^{\beta-2} \\
   x\,\varepsilon\,Q^{\alpha}\,\delta_{\alpha}^{\beta+2} &
   \nu^{1/2}\, Q^{-\alpha}\,\delta_{\alpha}^{\beta}\\
  \end{array}
\right),
\end{equation*}
with $\nu^{1/2}=(d_-/a_+)^{1/2}$, $b_-=c_+$,
$\varepsilon=b_-\,(a_+\,d_-)^{-1/2}$ and $\rho_{I}$ is an overall
factor constrained by the usual requirements of crossing and
unitarity. By performing a similarity transformation the $\alpha$
dependence of the off diagonal entries can be eliminated and the
type I defect matrix, as written in \cite{cz2010}, is recovered.
Then, the corresponding coefficients of the R matrix \eqref{TRT}
are:
\begin{eqnarray}\label{newboundarymatrixI}
&r_+=x\left(r\,q-\varepsilon^2{r^{-1} q^{-1}}\right)+{x^{-1}}
\left({r^{-1}}q-\varepsilon^2r{q^{-1}}\right),\nonumber\\
&r_-=x
\left(r^{-1}q-\varepsilon^2{r q^{-1}}\right) + {x^{-1}}
\left(r q-\varepsilon^2{r^{-1} q^{-1}}\right),\nonumber\\
&s_+=\left(x^2-x^{-2}\right)\varepsilon r^{-1}\nu^{-1/2},\nonumber\\
&s_-=\left(x^2-x^{-2}\right)\varepsilon r\nu^{1/2},
\end{eqnarray}
which is equivalent to the full Ghoshal-Zamolodchikov solution.

\p On the other hand, a `type II' defect is obtained  from
\eqref{Tmatrixg} by setting $a_+=d_-=1$, $a_-=-\bar b_+\,b_-$,
$d_+=-b_+\,\bar b_-\,q^{2}$, $c_+=-\bar b_-\,q^{2}$, $c_-=-\bar b_+$
and multiplying by $x$. Then the expression in \cite{cz2010} is
recovered. That is,
 \begin{equation*}
T_{II}\,^{b\, \beta}_{a \,\alpha}(\theta)=\rho_{II}(\theta)
\left(
\begin{array}{ccc}
  (Q^{\alpha}- b_-\, \bar b_+\, Q^{-\alpha}\, x^2)\,
  \delta^{\beta}_{\alpha} & x\,(b_+ Q^{\alpha} + b_- Q^{-\alpha})\,\delta^{\beta-2}_{\alpha} \\
  -x\,(\bar b_-\, Q^{\alpha-4} +\bar b_+\, Q^{-\alpha})
  \,\delta^{\beta+2}_{\alpha} & (-b_+\bar b_-\,Q^{\alpha-4}\, x^2+Q^{-\alpha} )\,\delta^{\beta}_{\alpha} \\
\end{array}
\right),
\end{equation*}
where the form of the scalar function $\rho_{II}$ is also
constrained by crossing and unitarity. The corresponding
coefficients of the reflection matrix \eqref{TRT} are:
\begin{eqnarray}\label{newboundarymatrixII}
&r_+=\left(x^2-x^{-2}\right)\,\left(b_+\bar b_-Q^{2\alpha-2}{r^{-1}}{x^{-1}}\,
- \bar b_+\,b_- Q^{-2\alpha-2}rx\right)
+x\,\left(gr-h{r^{-1}}\right)+
{x^{-1}}\left(g{r^{-1}}-h r\right),\nonumber\\
&r_-=\left(x^2-x^{-2}\right)\left(-b_+\bar b_-Q^{2\alpha-6}r^{-1}x
+\bar b_+ b_-Q^{-2\alpha+2}r{x^{-1}}\right)
+x\left(g{r^{-1}}-hr\right)+{x^{-1}}\left(gr-h{r^{-1}}\right),\nonumber\\
&s_+=\left(x^2-x^{-2}\right)\left(b_-{r^{-1}}+|b_+|^2 b_-r+b_+Q^{2\alpha}r^{-1}
+b_-^2\bar b_+Q^{-2\alpha}r\right),\nonumber\\
&s_-=\left(x^2-x^{-2}\right)\left(\bar b_- Q^{-4}r+|b_+|^2\bar b_-Q^{-4}{r^{-1}}
+\bar b_-^2 b_+Q^{2\alpha-8}r^{-1}+\bar b_+Q^{-2\alpha}r\right),
\end{eqnarray}
with
$$g=\left(|b_+|^2 |b_-|^2+1\right)Q^{-2},\qquad h=-\left(|b_-|^2 Q^{-2}+|b_+|^2\right)Q^{-2}.$$
These entries correspond to a new solution of the reflection
equation \eqref{bybe},  in which the explicit dependence on the
label $\alpha$ cannot be removed by a similarity transformation.

\p It is tempting to suggest, given the Lagrangian description of a
type II defect \cite{cz2009,cz2010}, that the system associated with
this reflection matrix is described classically (dropping explicit
reference to mass scale and the bulk coupling constant), by the
following Lagrangian density
\begin{equation}\label{boundaryL}
\mathcal{L}_B(u,\lambda)=\theta(-x)\,\mathcal{L}_u+\delta(x)(u\lambda_t-B(u,\lambda)),\qquad
\mathcal{L}_u=u_t^2/2 - u_x^2/2 - \left(e^u+e^{-u}\right),
\end{equation}
where $u$ is the sine/sinh-Gordon field (depending on whether $u$ is
imaginary or real), $\lambda$ is a time-dependent field defined at
the boundary and $B(u,\lambda)$ is a functional to be determined.
Forcing the energy-like spin three charge to be conserved (using the
argument appearing first in \cite{gz1994}), the form of the
functional $B$ is constrained to be:
\begin{equation}\label{Blambdadependence}
B(u,\lambda)=e^{\lambda/2}f(u)+e^{-\lambda/2}g(u),
\end{equation}
with
\begin{equation}\label{fg}
f(u)g(u)=h_+ e^{u/2}+h_- e^{-u/2}+2\,(e^u+e^{-u})+h_0,
\end{equation}
where $h_\pm$, $h_0$ are free constant parameters. Since  redefining
$\lambda \rightarrow \lambda+\Lambda(u)$ changes the Lagrangian by a
total derivative, it follows that the boundary part of the
Lagrangian density is essentially a three parameter expression,
which can be represented conveniently by \eqref{boundaryL} and
\eqref{Blambdadependence}  on choosing
\begin{equation}\label{fandg}
f(u)=f_0+\sqrt{2}(b e^{u/2}+b^{-1}e^{-u/2}),\qquad g(u)=g_0+\sqrt{2}(b^{-1} e^{u/2}+b e^{-u/2}).
\end{equation}
Notice that the most general solution of \eqref{fg} has an
additional  parameter. However, when the expressions of the
functionals $f$ and $g$ are used in \eqref{Blambdadependence} one
parameter can be eliminated by rescaling the field $\lambda$. It can
be verified that this expression for the boundary potential
corresponds to the expression obtained by adding together a type II
defect potential with the Ghoshal-Zamolodchikov Dirichlet boundary
potential (see \cite{bajnok07}, \cite{cz2010} for similar arguments).

\section{Discussion}

\p Some comments are in order at this point. A defect can be used to
create solutions to the reflection Yang-Baxter equation \eqref{bybe}
via \eqref{TRT}. Using the reflection matrix
\eqref{RmatrixDirichlet} representing any Dirichlet boundary
condition and the transmission matrix of a type I defect the
Ghoshal-Zamolodchikov solution \eqref{GZsolution} is recovered in
\eqref{newboundarymatrixI}. In this case the reflection matrix has
no dependence on topological charge other than via the Kronecker
deltas. Note, the simplest Dirichlet boundary condition imposes
$u(0,t)=0$, and its reflection factor has the form
\eqref{RmatrixDirichlet} with $r=1$; it is proportional to the
identity. This particular boundary condition  preserves the bulk
symmetry $u\rightarrow -u$ and, using a type I defect, the solution of
\eqref{bybe} is \eqref{newboundarymatrixI} with $r=1$. In this case,
the ensuing nonlinear boundary condition also maintains the bulk
symmetry $u\rightarrow -u$.

\p Also note, the results are consistent with an interesting
observation by Habibullin \cite{habibullin98} who pointed out how in
the classical sine-Gordon theory a general boundary condition could
be obtained from a general Dirichlet condition (that is, one with
$u_0\ne 0$), using a combination of two carefully designed
B\"acklund transformations. Explicitly, the combined effect of the
two B\"acklund transformations is effectively identical to the
Ghoshal-Zamolodchikov general boundary condition, namely,
$$u_x(0,t)=\kappa\sin \frac{(u+u_0)}{2} +\kappa^{-1}\sin\frac{(u-u_0)}{2},$$
where $\kappa$ is the parameter introduced by the B\"acklund
transformations.  It was pointed out also some time ago
\cite{bczlandau} that a type I defect is, from a classical point of
view, a B\"acklund transformation `frozen' at the location of the
defect. Therefore, the action of the defect in modifying a boundary
condition is essentially the Habibullin construction and \eqref{TRT}
is the quantum field theoretic version of this. Bajnok and Simon
\cite{bajnok07} noted the same phenomenon, albeit stated
differently, by adding a type I defect to a Dirichlet condition in
the sinh-Gordon model.

\p  Using a type II defect to generate solutions of \eqref{bybe}
leads inevitably to solutions that are outside the
Ghoshal-Zamolodchikov class since the explicit dependence of their
coefficients on topological charge is unavoidable. One may wonder
what boundary conditions might be responsible for these and one
suggestion has been proposed with the boundary Lagrangian
density \eqref{boundaryL}. The most general solution \eqref{cubicR}
to the transmission Yang-Baxter equation in the form \eqref{bybe}
has been derived explicitly and a comprehensive explanation
concerning the manner in which it is obtained using defects has been
provided. One could imagine that repeatedly placing type I or type
II defects near a Dirichlet boundary, or near a general
Ghoshal-Zamolodchikov boundary, should generate all possible
solutions to a (suitably generalised) version of the compatibility
relations \eqref{bybe} containing additional pairs of topological
charge labels (for example, see \eqref{TRT}). However, the types of
classical boundary condition to which they might effectively
correspond (if any) is unclear. One should bear in mind that any
process of reduction (or `fusing'), in which the additional sets of
labels collapse to a single set, must yield the already known general
solutions provided above, or a combination of them. On the other hand, the most general
solution to \eqref{STT} (with a single pair of topological charge
labels) is provided by \eqref{Tmatrixgg} (see also
\cite{Weston:2010cc} for an alternative algebraic argument). Hence
there is no other solution to \eqref{STT} that could be used in
combination with a Dirichlet boundary condition to provide a more
general solution to \eqref{bybe}.

\p It is worth recalling that
boundaries with additional degrees of freedom have been considered
previously for the sine-Gordon model by Baseilhac and Delius
\cite{DeliusBaseilhac01} and the corresponding reflection matrix
 found in \cite{BaseKoizumi03}. However, the latter represents a solution of a
boundary Yang-Baxter equation that does not preserve the topological
charge of the system. Hence it is not of the form \eqref{bybe}.
To make this fact more explicit, note that \eqref{Rmatrix} can be rewritten in terms of generalised annihilation and creation operators, $a,\, \hat a$ (as introduced in
\cite{Weston:2010cc} in the context of integrable defects) to find the alternative
\begin{equation}
\label{operatorRmatrix}
 R_a^b(\theta)=\left(\begin{array}{cc}
                                         r_+(N,x)& s_+(N,x)\, \hat a\\
                                         s_-(N,x)\, a&r_-(N,x)\\
                                        \end{array}\right),
\end{equation}
where $N$ is the generalised `number operator'. Thus,
$$a|j\rangle = |j-1\rangle,\ \hat a |j\rangle = F(j)|j+1\rangle,\ N|j\rangle=j|j\rangle, \ F(N)=a\hat a =f_0+f_1q^{2N}+f_2q^{-2N}.$$
Writing the expression in this manner makes it clear that diagonal elements satisfy
$$[r_\pm(N,x_1), r_\pm(N,x_2)]=0,$$
for any choices of sign, while the off-diagonal elements do not commute.
On the other hand, the solution provided for the reflection matrix in \cite{BaseKoizumi03}, in terms of the operators $\hat{\cal E}_\pm$, which, in the notation of that article, satisfy
\begin{equation}\label{BKalgebra}\left(q^2+q^{-2}\right)\hat{\cal E}_\pm \hat{\cal E}_\mp\hat{\cal E}_\pm -\hat{\cal  E}_\pm^2 \hat{\cal  E}_\mp-\hat{\cal  E}_\mp\hat{\cal  E}_\pm^2=-c^2\left(q+q^{-1}\right)^2\hat{\cal  E}_\mp,\end{equation}
has non-commuting diagonal elements leading to additional terms in the relations for the elements of the reflection matrix provided in the appendix\,\footnote{For example, the second equation of \eqref{twoterm} is replaced by the equation given as relation (i) on p501 of \cite{BaseKoizumi03}.}.

\p Finally, generalising to the complex affine Toda models,  the
analogue of the Dirichlet condition is given in
\eqref{todadirichlet} and it would be interesting to discover how
the reflection matrices corresponding to these are related (if
indeed they are) to the reflection matrices associated with the
boundary conditions \eqref{todaboundary}. The purpose of this note
was to emphasise the role defects might play and these specific
questions will be addressed in the future.

\vskip .5cm
\p {\large \bf Acknowledgements}\\

\p   One of us (EC) is grateful for the opportunity for discussions with Peter Bowcock and Craig Robertson and both of us wish to express our gratitude to the UK Engineering and Physical Sciences
Research Council for its support under grant reference EP/F026498/1. We are also grateful for the opportunity to work peacefully for a few weeks at CQUeST, Sogang University and YITP, Kyoto University,
within the framework provided by the European Commission grant FP7-PEOPLE-2009-IRSES-ISAQS.

\appendix
\section{Infinite dimensional solution to the reflection Yang-Baxter equation}
\label{appendixA}

Given the expression \eqref{Rmatrix} for the reflection matrix the reflection Yang-Baxter equation becomes a collection of nonlinear recurrence relations for the coefficients. These are of four basic types, 2-term, 4-term, 5-term and 6-term relations. It is also useful to define $a^\pm(x_1,x_2)=a(\theta_2\pm\theta_1),\ b^\pm(x_1,x_2)=b(\theta_2\pm\theta_1)$.

\p The 2-term relations involve just the off-diagonal elements and are:
\begin{eqnarray}\label{twoterm}
\nonumber &&s_\pm(\alpha,x_1)s_\pm(\alpha+2, x_2)=s_\pm(\alpha,x_2)s_\pm(\alpha+2, x_1),\\ &&s_+(\alpha,x_1)s_-(\alpha+2, x_2)=s_+(\alpha,x_2)s_-(\alpha+2, x_1),
\end{eqnarray}
implying $s_\pm(\alpha,x)=f(x) s_\pm(\alpha)$, where $f(x)$ depends on rapidity but not topological charge and the coefficients $s_\pm(\alpha)$ do not depend on rapidity.

\p There are a number of 4-term relations all of the following type,
\begin{eqnarray}\label{fourterm}
\nonumber &&s_+(\alpha,x_2)\left(a^-b^+\phantom{^1}r_+(\alpha+2 ,x_1)-a^+b^-\phantom{^1}r_+(\alpha,x_1)\right)\\
&&\phantom{mmmmmm}=c s_+(\alpha,x_1)\left(b^+ \phantom{^1}r_+(\alpha+2,x_2)-b^-\phantom{^1} r_-(\alpha+2,x_2)\right),
\end{eqnarray}
which simplify to
\begin{eqnarray}\label{fourterma}
\nonumber &&f(x_2)\left(a^-b^+\phantom{^1}r_+(\alpha+2 ,x_1)-a^+b^-\phantom{^1}r_+(\alpha,x_1)\right)\\
&&\phantom{mmmmmm}=c f(x_1)\left(b^+ \phantom{^1}r_+(\alpha+2,x_2)-b^-\phantom{^1} r_-(\alpha+2,x_2)\right),
\end{eqnarray}
and a number of 5-term relations of the form,
\begin{eqnarray}\label{fiveterm}
\nonumber &&s_+(\alpha,x_1)\left(a^+a^-\phantom{^1}r_+(\alpha ,x_2)-b^+b^-\phantom{^1}r_+(\alpha+2,x_2)-c^2r_-(\alpha+2,x_2)\right)\\
&&\phantom{mmmmmmmn}=c s_+(\alpha,x_2)\left(a^+\phantom{^1} r_+(\alpha,x_1)-a^- \phantom{^1}r_-(\alpha+2,x_1)\right).
\end{eqnarray}
Again, using \eqref{twoterm}, these simplify to,
\begin{eqnarray}\label{fiveterma}
\nonumber &&f(x_1)((a^-a^+\phantom{^1}r_+(\alpha ,x_2)-b^+b^-\phantom{^1}r_+(\alpha+2,x_2)-c^2\phantom{^1}r_-(\alpha+2,x_2))\\
&&\phantom{mmmmmmmn}=c f(x_2)\left(a^+ \phantom{^1}r_+(\alpha,x_1)-a^- \phantom{^1}r_-(\alpha+2,x_1)\right).
\end{eqnarray}
Finally, there is a 6-term relation that reads
\begin{eqnarray}\label{sixterm}
\nonumber &&cb^-(r_+(\alpha,x_1)r_+(\alpha,x_2)-r_-(\alpha,x_1)r_-(\alpha,x_2))\\
\nonumber&&\phantom{mmm}+c b^+(r_+(\alpha,x_1)r_-(\alpha,x_2)-r_+(\alpha,x_2)r_-(\alpha,x_1))\\
&&\phantom{mmmmmmm}+a^+b^-(s_+(\alpha,x_1)s_-(\alpha+2,x_2)-s_+(\alpha-2,x_1)s_-(\alpha,x_2))=0.
\end{eqnarray}
The latter is the only nonlinear relation to supply a real constraint on the off-diagonal functions of $\alpha$, $s_\pm(\alpha)$. Making the ansatz $f(x)\approx (x^2-1/x^2)$ and assuming $r_\pm$ are cubic in $x$ and $1/x$ leads (using Maple) to the solution given in \eqref{cubicR}. In effect equations \eqref{fourterma} and \eqref{fiveterma} are solved first for $r_\pm$ and the last equation \eqref{sixterm} strongly constrains $s_\pm(\alpha)$.

\p More generally, it is useful to consider the zeros of $f(x)$ and compare them with the zeros of $r_\pm(\alpha,x)$. Suppose $x_2$ is a root of $f(x)$ and $x_1$ is not a root of $f(x)$. Then, the right hand side of \eqref{fourterm} becomes
$$b^+ r_+(\alpha,x_2)-b^- r_-(\alpha,x_2)=0,$$
and therefore, using the explicit forms of $b^\pm$,
$$x_1^2\left[x_2^2\, r_+(\alpha, x_2)+r_-(\alpha, x_2)\right] -\left[x_2^2\, r_-(\alpha, x_2)+r_+(\alpha, x_2)\right]=0.$$
Thus, since this must hold for any choice of $x_1$,
$$x_2^2\,r_+(\alpha, x_2)+r_-(\alpha, x_2)=0,\quad x_2^2\, r_-(\alpha, x_2)+r_+(\alpha, x_2)=0,$$
which together imply $(1-x_2^4)r_\pm=0$. Thus, either $x_2$ is also a zero of $r_\pm(\alpha,x_2)$, or $x_2^4=1$. In other words, the only zeros of $f(x)$ not shared by the diagonal elements are located at the fourth roots of unity.

\end{document}